\documentstyle[preprint,aps]{revtex}
\tighten
\draft
\widetext
\input epsf
\preprint{YITP-SB-06-33}
\bigskip
\bigskip

\def\b{\overline}
\def\t{\widetilde}

\begin{document}
\title{A Novel Approach to the Cosmological Constant Problem}
\medskip
\author{
Alberto Iglesias$^1$\footnote{E-mail: iglesias@insti.physics.sunysb.edu} and
Zurab Kakushadze$^2$\footnote{E-mail: zurab.kakushadze@rbccm.com}}
\bigskip
\address{
$^1$C.N. Yang Institute for Theoretical Physics\\ 
State University of New York, Stony Brook, NY 11794\\
$^2$Royal Bank of Canada Dominion Securities Corporation\footnote{This address
is used by the corresponding author for no purpose other than to indicate his
professional affiliation as is customary in scientific publications. In 
particular, the contents of this paper are limited to Theoretical Physics, have
no relation whatsoever to the Securities Industry, do not have any value for
making investment or any other financial decisions, and in no way represent
views of the Royal Bank of Canada or any of its other affiliates.}\\
1 Liberty Plaza, 165 Broadway, New York, NY 10006}
\date{June 30, 2003}
\bigskip
\medskip
\maketitle

\begin{abstract} 
{}We propose a novel infinite-volume brane world scenario where we live 
on a non-inflating spherical 3-brane, whose radius is
somewhat larger than the present Hubble size, embedded in higher 
dimensional bulk. Once we include higher 
curvature terms in the bulk, we find completely smooth solutions with 
the property that the 3-brane world-volume is non-inflating for a continuous
range of positive values of the brane tension, that is, without fine-tuning.
In particular, our solution, which is a near-BPS background with
supersymmetry broken on the brane around TeV, is controlled by a single
integration constant.
\end{abstract}
\pacs{}

\section{Introduction and Summary}

{}As was proposed in \cite{DGP,DG}, one can reproduce 4-dimensional gravity
on a 3-brane embedded in higher dimensional infinite-volume bulk if one
includes the Einstein-Hilbert term on the brane. Gravity becomes higher
dimensional above some cross-over scale, which can be larger than the present
Hubble size. As was pointed out in \cite{DGP,witten,DG}, infinite-volume
scenarios offer a new arena for addressing the cosmological constant problem.
In particular, since at large distances gravity is no longer 4-dimensional, it
is possible that the usual four-dimensional arguments for why fine-tuning is
unavoidable are no longer valid. What one needs in this context, however, 
is to construct explicit solutions where the 3-brane tension is large yet the
4-dimensional cosmological constant is vanishing (or small) without 
fine-tuning.  

{}The goal is to construct infinite-volume solutions where the
brane is non-inflating (or very slowly inflating) with the 
property that such solutions exist for a continuous range of positive
values of the 3-brane tension. One approach is to look for solutions
where the 3-brane world-volume is flat. 
Such solutions do not exist for codimension-1 branes \cite{DGP}. 
In the codimension-2 case such solutions do exist, but the brane tension 
has an upper bound such that it does not improve the experimental bound 
on the 4-dimensional cosmological constant \cite{CIKL}. One is therefore 
compelled to go to codimension-3 and higher setups. 

{}Here one encounters the presence of naked singularities \cite{Gregory} 
unless one includes higher curvature terms in the bulk action 
\cite{Higher,CCP}. However, inclusion of higher curvature terms in the
bulk does indeed cure the singularity problem. 
Thus, \cite{CCP} constructed explicit 
{\em non-singular} higher codimension solutions with the property that
the brane world-volume contains the 4-dimensional Minkowski 
space\footnote{Since
non-inflating smooth solutions constructed in \cite{CCP} do not require
fine-tuning, we expect that consistent slowly inflating solutions should also
exist (explicitly constructing such solutions is, however, technically
challenging) without fine-tuning. 
In \cite{DGS} it was proposed that considering slowly inflating
solutions {\em without} higher curvature terms in the bulk might solve the 
singularity problem. It would be interesting to find explicit solutions
with this property.}, and such
solutions exist for a continuous range of positive values of the brane 
tension. That is, the solutions of \cite{CCP} provide 
explicit examples of infinite-volume
brane world solutions where vanishing of the brane cosmological constant 
does not require fine-tuning.
One of the key ingredients in \cite{CCP} is that
the brane in these solutions is not just a flat 3-brane. Rather, 
following the proposal of \cite{orient}, the brane geometry is that of
the 4-dimensional Minkowski space times a small $n$-sphere ($n\ge 2$). 

{}In this paper we take various observations of \cite{CCP} even further. Our 
main idea here is that {\em a priori} there is no reason why the 4-dimensional
part of the brane world-volume should be flat. Thus, consider a 3-brane
whose world-volume is a 3-sphere of a huge radius (somewhat larger than the
present Hubble size). This is perfectly consistent with observation. We can
then imagine a scenario where this brane is embedded in, say, 5-dimensional 
bulk\footnote{Higher
dimensional generalizations can also be considered - see below.}. To
avoid singularities we must include higher curvature terms in the bulk action.
Now we can ask whether we can find non-singular solutions where the 3-brane
is non-inflating for a continuous range of large positive values of the brane
tension. This resembles the setup of \cite{CCP}, but the key difference
here is that the 3-brane itself is a sphere (and the only flat direction 
is time) of a huge radius.

{}In this paper we present {\em explicit non-singular} solutions of this 
type. To make the problem computationally tractable, we choose the higher
curvature terms in the bulk to be the quadratic (in curvature)
Gauss-Bonnet combination\footnote{In fact, when the higher curvature terms are
of the Gauss-Bonnet type, the problem turns out
to be solvable {\em analytically}, so all of the following statements can be 
verified via explicit calculations.} (we expect that such solutions should
exist for generic higher curvature terms as
well, albeit we do not have a proof of this statement). 
In our solutions, which are spherically symmetric, the space is asymptotically
flat in both $r\rightarrow 0$ and $r\rightarrow +\infty$ limits, but it 
is curved for finite $r$. There are no 
singularities whatsoever, the volume of the extra space is infinite, and
the brane tension takes values
in a continuous range, so we do not need any fine-tuning. In particular, 
our solution, which is a near-BPS background with
supersymmetry broken on the brane around TeV, is controlled by a single
integration constant.

{}Finally, let us point out that, in the 5-dimensional setup we focus on in
this paper, the bulk Planck scale should be of order of the 4-dimensional
Planck scale to have: supersymmetry breaking around TeV; the radius of the
3-brane above the present Hubble size; no need for fine-tuning the brane
tension. Then, even in the presence of the Einstein-Hilbert 
term on the brane, gravity on the brane is actually 5-dimensional. This 
is because there is no hierarchy between the bulk and brane Planck scales
while the volume of the extra dimension is infinite. One 
approach to circumvent this, which we comment on in the last section, 
would be to consider higher dimensional setups where
a spherical 3-brane is embedded in higher-than-five dimensional bulk.
At any rate, our results suggest that it might be worthwhile to study
string theory backgrounds in the presence of large closed (spherical)
D3-branes (and, more generally, $p$-branes).

\section{Setup}

{}The brane world model we study in this paper is described 
by the following action:
\begin{eqnarray}
 S=&&{\widetilde M}^{D-3}_P \int_{\Sigma} d^{D-1}x~
 \sqrt{-{\widetilde G}}\left[{\widetilde R}-{\widetilde\Lambda}\right]
 +\nonumber\\
 &&{M}^{D-2}_P \int d^{D}x~
 \sqrt{-{G}}~\left[{R}~+\xi\biggl(R^2-4R_{MN}^2+R_{MNPR}^2\biggl)\right].
 \label{action}
\end{eqnarray}
Here $M_P$ is the (reduced) $D-$dimensional Planck mass, while 
${\widetilde M}_P$
is the (reduced) $(D-1)$-dimensional Planck mass; $\Sigma$ is a 
source brane, whose geometry is given by 
the product ${\bf R}^{D-d-1,1}\times {\bf S}^{d-1}_\epsilon$, where 
${\bf R}^{D-d-1,1}$ is the $(D-d)$-dimensional Minkowski space, and 
${\bf S}^{d-1}_\epsilon$ is a $(d-1)$-sphere of radius $\epsilon$ (in the
following we will assume that $d\geq 3$).
The quantity ${\widetilde M}^{D-1}_P {\widetilde \Lambda}$ plays the role of 
the tension of the brane $\Sigma$. Also, 
\begin{equation}
 {\widetilde G}_{mn}\equiv
 {\delta_m}^M{\delta_n}^N G_{MN}\Big|_\Sigma~,
\end{equation} 
where $x^m$ are the $(D-1)$ coordinates along the brane (the $D$-dimensional
coordinates are given by $x^M=(x^m,r)$, where $r\geq 0$ is a non-compact
radial coordinate transverse to the brane, and the signature of the 
$D$-dimensional metric is $(-,+,\dots,+)$); finally, the $(D-1)$-dimensional
Ricci scalar ${\widetilde R}$ is constructed from the $(D-1)$-dimensional
metric ${\widetilde G}_{\mu\nu}$. In the following we will use the notation
$x^i=(x^\alpha,r)$, where $x^\alpha$ are the $(d-1)$ angular 
coordinates on the sphere ${\bf S}^{d-1}_\epsilon$.
Moreover, the metric for the coordinates $x^i$ will be (conformally) flat:
\begin{equation}
 \delta_{ij}~dx^i dx^j=dr^2+r^2\gamma_{\alpha\beta}~dx^\alpha dx^\beta~,
\end{equation}
where $\gamma_{\alpha\beta}$ is the metric on a unit $(d-1)$-sphere.
Also, we will
denote the $(D-d)$ Minkowski coordinates on ${\bf R}^{D-d-1,1}$ via $x^\mu$
(note that $x^m=(x^\mu,x^\alpha)$).

{}The equations of motion read
\begin{eqnarray}
 &&R_{MN}-\frac{1}{2} G_{MN}R+\xi \left[2\Xi_{MN}-{1\over 2} G_{MN}\, 
 \Xi\right]+
 \nonumber\\
 &&\displaystyle{
 {\sqrt{-{\widetilde G}}\over\sqrt{-G}}}
 {\delta_M}^m {\delta_N}^n \left[{\widetilde R}_{mn}-{1\over 2} 
 {\widetilde G}_{mn}\left({\widetilde R}-
{\widetilde \Lambda}\right)\right]  {\widetilde L}~ \delta(r-\epsilon)=0~,
 \label{EoM4}
\end{eqnarray}
where $\Xi\equiv {\Xi_M}^M$, and
\begin{eqnarray}
 &&\Xi_{MN}\equiv 
 R R_{MN}-2R_{MS}R^S{}_N+R_{MRST}R_N{}^{RST}-2R^{RS}R_{MRNS}~,\\
 &&{\widetilde L}\equiv {\widetilde M}_P^{D-3}/M_P^{D-2}~.
\end{eqnarray}
Here we are interested in solutions with vanishing $(D-d)$-dimensional
cosmological constant, which, at the same time, are radially symmetric in the
extra $d$ dimensions. The corresponding ansatz for the background metric reads:
\begin{equation}
 ds^2=\exp(2A)~\eta_{\mu\nu}~dx^\mu dx^\nu+\exp(2B)~\delta_{ij}~dx^i dx^j~,
\end{equation}
where $A$ and $B$ are
functions of $r$ but are independent of $x^\mu$ and $x^\alpha$. 
We then have (here prime
denotes derivative w.r.t. $r$):
\begin{eqnarray}
 &&{\widetilde R}_{\mu\nu}=0~,\\
 &&{\widetilde R}_{\alpha\beta}=\lambda ~{\widetilde G}_{\alpha\beta}~,\\
 &&{\widetilde R}=(d-1)~\lambda~,\\
 &&R_{\mu\nu}=-\eta_{\mu\nu}e^{2(A-B)}
 \left[A^{\prime\prime} +(d-1){1\over r} A^\prime+
 (D-d)(A^\prime)^2 +(d-2)A^\prime B^\prime\right]~,\\
 &&R_{rr}=-(d-1)\left[B^{\prime\prime}+ {1\over r} B^\prime\right]
 +(D-d)\left[A^\prime B^\prime-
 (A^\prime)^2 -A^{\prime\prime}\right]~,\\
 &&R_{\alpha\beta}=-r^2 \gamma_{\alpha\beta}
 \left[B^{\prime\prime}+(2d-3){1\over r}B^\prime +(d-2)(B^\prime)^2+
 (D-d)A^\prime B^\prime + (D-d){1\over r}A^\prime\right]~,\\
 &&R=-e^{-2B}\Big[2(d-1)B^{\prime\prime} + 2(d-1)^2{1\over r} B^\prime +
 2(D-d)A^{\prime\prime} + 2(D-d)(d-1){1\over r} A^\prime 
 +\nonumber\\
 &&(d-1)(d-2)(B^\prime)^2 +(D-d)(D-d+1)(A^\prime)^2
 +2(D-d)(d-2)A^\prime B^\prime\Big]~. 
\end{eqnarray}
where
\begin{equation}
 \lambda\equiv {{d-2}\over \epsilon^2}e^{-2B(\epsilon)}~.
\end{equation}
The equations of motion then read:
\begin{eqnarray}\label{AB1}
 &&{1\over 2}\b d (\b d-1)(A^\prime)^2+\b d(d-1)N+{1\over 2}(d-1)(d-2)S-
 \nonumber\\
 &&\xi\ {\rm e}^{-2B}\biggl\{{1\over 2}\b d(\b d -1)(\b d-2)(\b d-3)
 (A^\prime)^4
 +2\b d(\b d -1)(d-1)(d-2)N^2+\nonumber\\
 &&{1\over 2}(d-1)(d-2)(d-3)(d-4)S^2
 +2\b d(\b d -1)(\b d-2)(d-1)(A^\prime)^2N
 \nonumber\\
 &&+\b d(\b d -1)(d-1)(d-2)(A^\prime)^2S
 +2\b d(d-1)(d-2)(d-3)NS\biggr\}=0~,\\%[3mm]
 &&{1\over 2}(\b d-1)(\b d-2)(A^\prime)^2+(\b d-1)M+(\b d-1)(d-1)N
+(d-1)P+{1\over 2}(d-1)(d-2)S-\nonumber\\
&&\xi\ {\rm e}^{-2B}\biggl\{{1\over 2}(\b d-1)(\b d-2)(\b d-3)(\b d-4)
(A^\prime)^4+2(\b d-1)(\b d -2)(d-1)(d-2)N^2+
\nonumber\\&&{1\over 2}(d-1)(d-2)(d-3)(d-4)S^2
+2(\b d-1)(\b d-2)(\b d-3)(A^\prime)^2M+
\nonumber\\&&2(\b d-1)(\b d -2)(\b d-3)(d-1)(A^\prime)^2N
+2(\b d-1)(\b d-2)(d-1)(A^\prime)^2P+
\nonumber\\&&(\b d-1)(\b d-2)(d-1)(d-2)(A^\prime)^2S
+4(\b d-1)(\b d-2)(d-1)MN+\nonumber\\
&&2(\b d-1)(d-1)(d-2)MS+4(\b d-1)(d-1)(d-2)NP+\nonumber\\
&&2(\b d-1)(d-1)(d-2)(d-3)NS+2(d-1)(d-2)(d-3)PS\biggr\}+
\nonumber\\&&{1\over 2}{\rm e}^{B}
\biggl[\t\Lambda-(d-1)\lambda\biggr]\t L\delta(r-\epsilon)=0~,\label{AB2}\\
%[3mm]
%
%
%
&&{1\over 2}\b d(\b d-1)(A^\prime)^2+\b dM+\b d(d-2)N
+(d-2)P+{1\over 2}(d-2)(d-3)S-\nonumber\\
&&\xi\ {\rm e}^{-2B}\biggl\{{1\over 2}\b d(\b d-1)(\b d-2)(\b d-3)
(A^\prime)^4+2\b d(\b d -1)(d-2)(d-3)N^2+
\nonumber\\&&{1\over 2}(d-2)(d-3)(d-4)(d-5)S^2
+2\b d(\b d-1)(\b d-2)(A^\prime)^2M+
\nonumber\\&&2\b d(\b d-1)(\b d -2)(d-2)(A^\prime)^2N
+2\b d(\b d-1)(d-2)(A^\prime)^2P+
\nonumber\\&&\b d(\b d-1)(d-2)(d-3)(A^\prime)^2S
+4\b d(\b d-1)(d-2)MN+2\b d(d-2)(d-3)MS+
\nonumber\\&&4\b d(d-2)(d-3)NP
+2\b d(d-2)(d-3)(d-4)NS+2(d-2)(d-3)(d-4)PS\biggr\}+
\nonumber\\&&{1\over 2}{\rm e}^{B}
\biggl[\t\Lambda-(d-3)\lambda\biggr]\t L\delta(r-\epsilon)=0~,\label{AB3}
\end{eqnarray}
where we have defined
\begin{eqnarray}
&&M\equiv A^{\prime\prime}+(A^\prime)^2-A^\prime B^\prime~,\hskip1cm
N\equiv A^\prime \left(B^\prime +{1\over r}\right)~,\nonumber\\
&&P\equiv B^{\prime\prime}+{B^\prime\over r}~,\hskip1cm
S\equiv (B^\prime)^2+2{B^\prime\over r}~,\nonumber\\ 
&&\b d \equiv D-d~.\nonumber
\end{eqnarray}
Above the third equation is the $(\alpha\beta)$ equation, 
the second equation is the
$(\mu\nu)$ equation, while the first equation is the $(rr)$ equation.
Note that the latter 
does not contain second derivatives of $A$ and $B$. Also note that the other
two equations do not contain third and fourth derivatives of $A$ and $B$
(this is a special property of the Gauss-Bonnet combination).

{}The equations of motion (\ref{AB1})-(\ref{AB3}) are highly non-linear and
difficult to solve in the general case. However, in this paper we focus on
the case ${\overline d}=1$ and $d=4$. That is, our brane has the geometry of
${\bf R}^{0,1} 
\times {\bf S}^3$ (the first factor corresponds to the time direction),
and it is embedded in a 5-dimensional bulk. As we will see in the next
section, we can then solve these equations analytically.

\section{Spherical 3-brane in 5D Bulk}

{}In the ${\overline d}=1$ and $d=4$ case the equations of motion
simplify substantially:
\begin{eqnarray}
 \label{firstder}
 &&N+S-4\xi {\rm e}^{-2B}NS=0~,\\
 \label{onlyB}
 &&P+S-4\xi {\rm e}^{-2B}PS+{1\over 6}{\rm e}^B\left[{\widetilde\Lambda}-
 3\lambda\right]{\widetilde L}\delta(r-\epsilon)=0~,\\
 \label{mixed}
 &&M+2N+2P+S-4\xi {\rm e}^{-2B}\left[MS+2NP\right]+
 {1\over 2}{\rm e}^B\left[{\widetilde\Lambda}-\lambda\right]{\widetilde L}
 \delta(r-\epsilon)=0~. 
\end{eqnarray}
Note that $P$ and $S$ do not contain $A$, so the second of the above equations
is a (non-linear) ordinary differential equation for $B$. We can actually
solve it analytically.

{}To begin with, we will solve this equation without the $\delta$-function
source term. Then we will find appropriate solutions by imposing the
matching conditions. Let us introduce the following simplifying notations:
\begin{eqnarray}
 &&r_0^2\equiv 4\xi~,\\
 &&C\equiv B+\ln(r/r_0)~,\\
 &&z\equiv \ln(r/r_*)~,
\end{eqnarray}
where $r_*$ is an arbitrary positive parameter (which, as will become clear
in the following, is just an integration constant). Then (\ref{onlyB}) reads
(without the source term):
\begin{equation}
 C_{zz}={{1-(C_z)^2}\over{1+\left[1-(C_z)^2\right]{\rm e}^{-2C}}}~,
\end{equation}
where subscript $z$ denotes derivative w.r.t. $z$. This equation can be
solved as follows. Let
\begin{equation}
 1-(C_z)^2\equiv (f(C)-1){\rm e}^{2C}~,
\end{equation}
where $f(C)$ is a function of $C$ only. Then we have the following first
order differential equation for $f(C)$:
\begin{equation}
 f_C=-2(f^2-1)/f~,
\end{equation}
whose solution is given by
\begin{equation}
 f(C)=\eta\sqrt{1+\omega {\rm e}^{-4C}}~,
\end{equation}
where $\eta=\pm 1$ corresponds to two distinct branches, and $\omega$ is the
integration constant ($\omega\in {\bf R}$).
The solution for $C$ is then
given by:
\begin{equation}
 C(z)=F(\zeta z,\eta,\omega)~,
\end{equation}
where 
$F(x,\eta,\omega)$ is a solution of the following differential
equation (note that this solution depends on another integration constant)
\begin{equation}
 {dF\over dx}=\sqrt{1+{\rm e}^{2F}-\eta\sqrt{\omega+{\rm e}^{4F}}}~,
\end{equation} 
and $\zeta=\pm 1$ corresponds to two distinct branches. 

\subsection{Non-singular Solutions without the Source Brane}

{}Here we would like to see if there are {\em non-singular} solutions
(in the absence of the source terms - if we find such solutions, we can then
see if we can construct solutions with the source terms with desirable
properties). Such a solution indeed exists. Thus, consider the case
where $\eta=-1$, $\zeta=+1$ and $\omega=0$. We then have
(in this solution $-\infty<z<z_1$):
\begin{equation}
 C(z)=-\ln\left[-\sqrt{2} \sinh\left(z-z_1\right)\right]~,
\end{equation}
where $z_1$ is the remaining integration constant (note that we have set
the first integration constant $\omega=0$). From (\ref{firstder}) 
it follows that
\begin{equation}\label{A}
 A=\ln\left|C_z\right|+{\rm constant}~.
\end{equation}
This implies that in the above solution we have
\begin{equation}
 A(z)=-\ln\left[-\tanh\left(z-z_1\right)\right]+{\rm constant}~. 
\end{equation}
The asymptotic behavior of this solution is given by:
\begin{eqnarray}
 &&C(z\rightarrow -\infty)=z+{\cal O}(1)~,\\
 &&A(z\rightarrow -\infty)={\rm constant}~,\\
 &&C(z\rightarrow z_1 -)=-\ln\left(z_1-z\right)+{\cal O}(1)~,\\
 &&A(z\rightarrow z_1 -)=-\ln\left(z_1-z\right)+{\cal O}(1)~.
\end{eqnarray}
Thus, when $r\rightarrow 0$ we have
\begin{eqnarray}
 &&B(r\rightarrow 0)={\rm constant}~,\\
 &&A(r\rightarrow 0)={\rm constant}~,
\end{eqnarray}
that is, in this limit the space is asymptotically flat.

{}What about $z\rightarrow z_1-$ limit? The Ricci scalar is given by:
\begin{equation}
 R=-{2\over r_0^2}{\rm e}^{-2C}\left\{3\left[C_{zz}+(C_z)^2-1\right]+
 A_{zz}+2A_z C_z +(A_z)^2\right\}~.
\end{equation}
In the above limit the Ricci scalar actually goes to a constant:
\begin{equation}
 R\rightarrow -{40\over r_0^2}~.
\end{equation}
Note, however, that the space is {\em not} asymptotically AdS$_5$ in this
limit (in particular, our solution is spherically symmetric, while AdS$_5$ is 
not). Moreover, consider the volume outside of a 3-sphere of radius
$b$ (where $b<r_1\equiv r_* {\rm e}^{z_1}$):
\begin{equation}
 v(b)=v_3 \int_b^{r_1} dr~r^3~ {\rm e}^{A+4B}=v_3~r_0^4 
 \int_{z_b}^{z_1} dz~{\em e}^{A+4C}~,
\end{equation}
where $v_3$ is the volume of a unit 3-sphere, and $z_b\equiv\ln(b/r_*)<z_1$.
The volume $v(b)$ is actually {\rm infinite}. Thus, as $z\rightarrow z_1-$
we have a coordinate singularity (and not a true singularity). In particular,
all geodesics are complete in this limit. 

{}Finally, let us mention that we also have the $\zeta=-1$ counterpart of the
above solution, that is, the solution with $\eta=-1$, $\zeta=-1$ and 
$\omega=0$, which is also non-singular. This solution is given by (in this
solution $z_2<z<+\infty$):
\begin{eqnarray}
 &&C(z)=-\ln\left[\sqrt{2}\sinh\left(z-z_2\right)\right]~,\\
 &&A(z)=-\ln\left[\tanh\left(z-z_2\right)\right]+{\rm constant}~,
\end{eqnarray}
where $z_2$ is an integration constant. In constructing the full solution
with the source term we will use the $\zeta=+1$ solution. Also, in the
full solution we will not even have a coordinate singularity (as the brane
will be located to the left of this would-be coordinate singularity, while
to the right of the brane we will have another solution, so that the resulting
full solution is completely non-singular).

\subsection{Matching Conditions}

{}Note that (\ref{firstder}) contains only first derivatives of $A$ and $B$, 
and it does not contain a $\delta$-function source term. This equation,
therefore, is satisfied by the above solutions. However, the other
two equations (\ref{onlyB}) and (\ref{mixed}) contain second derivatives and
source terms, and produce non-trivial matching conditions. We must therefore
make sure that they are satisfied. In the $z$-coordinate frame these three
equations read:
\begin{eqnarray}
 \label{first-z-der}
 &&A_z C_z\left\{1+\left[1-(C_z)^2\right]{\rm e}^{-2C}\right\}=
 1-(C_z)^2~,\\
 \label{onlyC}
 &&C_{zz}\left\{1+\left[1-(C_z)^2\right]{\rm e}^{-2C}\right\}=
 \left[1-(C_z)^2\right]-{1\over 6}{\rm e}^C\left[{\widetilde\Lambda}-
 3\lambda\right]r_0{\widetilde L}\delta(z-z_\epsilon)~,\\
 \label{AandC}
 &&\left[A_{zz}+(A_z)^2\right]
 \left\{1+\left[1-(C_z)^2\right]{\rm e}^{-2C}\right\} +
 2C_{zz}\left\{1-A_z C_z {\rm e}^{-2C}\right\}=\nonumber\\
 &&\phantom{}~~~~~~2\left[1-(C_z)^2-A_z C_z\right]-
 {1\over 2}{\rm e}^C\left[{\widetilde\Lambda}-
 \lambda\right]r_0{\widetilde L}\delta(z-z_\epsilon)~,
\end{eqnarray}
where 
\begin{equation}
 z_\epsilon\equiv\ln(\epsilon/r_*)~,
\end{equation}
and in arriving at (\ref{AandC}) we have used (\ref{firstder}).

{}The third equation (\ref{AandC}) looks cumbersome. We can simplify it as
follows. Note that $C_z$ is always non-vanishing. We can therefore rewrite
(\ref{first-z-der}) as follows:
\begin{equation}\label{A_z}
 A_z \left\{1+\left[1-(C_z)^2\right]{\rm e}^{-2C}\right\}+C_z-1/C_z=0~.
\end{equation}
Differentiating this equation w.r.t. $z$ we obtain:
\begin{equation}
 A_{zz} \left\{1+\left[1-(C_z)^2\right]{\rm e}^{-2C}\right\}-
 2A_z C_z\left\{C_{zz}+\left[1-(C_z)^2\right]\right\}{\rm e}^{-2C} +
 C_{zz}\left[1+1/(C_z)^2\right]=0~.
\end{equation}
Equation (\ref{AandC}) can then be rewritten as follows:
\begin{equation}\label{AC}
 \left[C_{zz}-A_z C_z\right]
 \left[1-1/(C_z)^2\right]=
 -{1\over 2}{\rm e}^C\left[{\widetilde\Lambda}-
 \lambda\right]r_0{\widetilde L}\delta(z-z_\epsilon)~,
\end{equation}
which is much nicer than the original equation (\ref{AandC}).

{}Note that (\ref{AC}) gives (\ref{A}) away from the brane, and we 
obtained (\ref{A}) away from the brane from the first two equations
(\ref{firstder}) and (\ref{onlyB}). This is, as usual, a consequence of
the diffeomorphism invariance - away from the brane only two out of the
three equations are independent (which is just as well for we have only two
warp factors $A$ and $B$). However, since the brane is a source brane (as 
opposed to a solitonic one), at the brane the diffeomorphism invariance is
not all intact (more precisely, it is partially reduced by an implicit gauge
choice). As a result, (\ref{AC}) does produce a non-trivial matching condition
at the brane (that is, at $z=z_\epsilon$), and so does (\ref{onlyC}).
 
{}Thus, we have the following two matching conditions from (\ref{onlyC}) and
(\ref{A}), respectively:
\begin{eqnarray}
 && \left. C_z\left\{1+\left[1-{1\over 3}(C_z)^2\right]{\rm e}^{-2C}\right\}
 \right|^{z_\epsilon +}_{z_\epsilon -}=
 -{1\over 6}{\rm e}^{C(z_\epsilon)}r_0{\widetilde L}\left[{\widetilde 
 \Lambda}-3\lambda\right]~,\\
 && \left. \left[C_z+1/C_z\right]\right|^{z_\epsilon +}_{z_\epsilon -}=
 -{1\over 2}{\rm e}^{C(z_\epsilon)}r_0{\widetilde L}\left[{\widetilde 
 \Lambda}-\lambda\right]~.
\end{eqnarray}
Let us simplify these conditions as follows. First, let
\begin{equation}
 \kappa={\widetilde \Lambda}/\lambda~.
\end{equation}
In the following we will set the corresponding integration constant so
that
\begin{equation}
 A(z=z_\epsilon)=0~.
\end{equation}
Next, note that the radius of the 3-sphere actually is not $\epsilon$, 
rather, it is given by
\begin{equation}
 {\widetilde R}\equiv
 \epsilon ~{\rm e}^{B(r=\epsilon)}=r_0 {\rm e}^{C(z_\epsilon)}~.
\end{equation}
Finally, recall that
\begin{equation}
 \lambda={2\over\epsilon^2}{\rm e}^{-2B(r=\epsilon)}={2\over 
 {\widetilde R}^2}~.
\end{equation}
With these notations we then have:
\begin{eqnarray}\label{match1}
 && \left. C_z\left\{1+{r_0^2\over 
 {\widetilde R}^2}\left[1-{1\over 3}(C_z)^2\right]\right\}
 \right|^{z_\epsilon +}_{z_\epsilon -}=
 -{{\kappa-3}\over 3}{{\widetilde L}\over {\widetilde R}}~,\\
 \label{match2}
 && \left. \left[C_z+1/C_z\right]\right|^{z_\epsilon +}_{z_\epsilon -}=
 -(\kappa-1){{\widetilde L}\over {\widetilde R}}~.
\end{eqnarray}
What we would like to see is if these two equations have a solution for
$C_z$ and $\kappa$. More precisely, we need two things. First, to
successfully address the cosmological constant problem, solutions must
exist for a continuous range of values of $\kappa$ (or else ${\widetilde
\Lambda}$ would have to be fine tuned). This is actually guaranteed if we find
one solution - indeed, the size of the 3-sphere 
${\widetilde R}$ is not a parameter of the
theory, it contains a free integration constant via $C(z_\epsilon)$ (it is
clear that $\kappa$ does not scale exactly as ${\widetilde R}^2$). 
Second, we must make sure that solutions exist for a phenomenologically 
interesting range of parameters. In the next subsection we give explicit
solutions of the aforementioned type.

\subsection{Explicit Non-Singular Solutions with the Source Brane}

{}We are interested in constructing a solution such that the space is
at least asymptotically
flat for $r>\epsilon$, and the volume in the extra dimension is infinite.
With the aforementioned phenomenological constraints, we are then led to
consider a solution of the following type (we have chosen the integration
constant in $A(z)$ such that $A(z_\epsilon)=0$; moreover, we have absorbed
the integration constant $z_1$ from above into the definition of $r_*$):
\begin{eqnarray}
 z>z_\epsilon:~~~&&C(z)=\Phi(z,\omega)~,\\
		 &&A(z)=\ln\left({\Phi_z(z,\omega)\over
 \Phi_z(z_\epsilon,\omega)}\right)~,\\
 z<z_\epsilon:~~~&&C(z)=-\ln\left[-\sqrt{2}\sinh(z)\right]~,\\
		 &&A(z)=-\ln\left[{\tanh(z)\over \tanh(z_\epsilon)}\right]~,
\end{eqnarray}
where $\Phi(z,\omega)$ is the solution of the following differential 
equation 
\begin{equation}
 \Phi_z=\sqrt{1+{\rm e}^{2\Phi}-\sqrt{\omega+{\rm e}^{4\Phi}}}
\end{equation}
subject to the initial condition
\begin{equation}
 \Phi(z_\epsilon,\omega)=-\ln\left[-\sqrt{2}\sinh(z_\epsilon)\right]~.
\end{equation}
Here $\omega$ is an integration constant. It is, however, constrained as
follows (with this constraint the above full
solution is non-singular for $z\in{\bf R}$):
\begin{equation}
 -{1\over 4\sinh^4(z_\epsilon)}<\omega<\coth^2(z_\epsilon)~.
\end{equation}
Note that $\epsilon<r_*$ so that $z_\epsilon<0$. Moreover, the solution is
asymptotically flat in both $z\rightarrow\pm\infty$ limits, and the volume
of the fifth dimension is {\em infinite}.

{}In the following it will be more convenient to parametrize $\omega$ as 
follows:
\begin{equation}
 \omega\equiv \nu\coth^2(z_\epsilon)~,
\end{equation}
where
\begin{equation}
 -{1\over \sinh^2(2z_\epsilon)}<\nu<1~.
\end{equation}
Note that
\begin{equation}
 -\sinh(z_\epsilon)={r_0\over \sqrt{2}{\widetilde R}}\ll 1~.
\end{equation}
That is, $-z_\epsilon\ll 1$, and 
the brane is located very close to the would-be coordinate 
singularity (at $r=r_*$):
\begin{equation}
 {\epsilon\over r_*}-1\approx -{r_0\over \sqrt{2}{\widetilde R}}~.
\end{equation}
We then have (for the reasons that will become clear in the following, here
we are assuming $|\nu|\sim 1$):
\begin{eqnarray}
 &&C_z(z_\epsilon -)= -1/z_\epsilon+{\cal O}(z_\epsilon)~,\\
 &&C_z(z_\epsilon +)= \sqrt{1-\nu} +{\cal O}(z_\epsilon^2)~.
\end{eqnarray}
The matching conditions (\ref{match1}) and (\ref{match2}) 
then read:
\begin{eqnarray}
 &&\sqrt{2}{{\widetilde L}\over r_0}=
 {1\over z_\epsilon}{{2\nu-1}\over 2\sqrt{1-\nu}}-{3\over 2}+
 {\cal O}(z_\epsilon)~,\\
 &&\sqrt{2}\kappa{{\widetilde L}\over r_0}={1\over z_\epsilon^2}+
 {1\over z_\epsilon}{3\over 2\sqrt{1-\nu}}+
 {\cal O}(1)~.
\end{eqnarray}
Let
\begin{equation}
 {\widetilde \nu}\equiv {1\over 2}-\nu~.
\end{equation}
Then we have (the fact that ${\widetilde\nu}$ must be small will become
clear momentarily):
\begin{eqnarray}
 &&0<{\widetilde \nu}\ll 1~,\\
 &&z_\epsilon\approx -{\widetilde \nu} 
 \left[{{\widetilde L}\over r_0}+{3\over 2\sqrt{2}}\right]^{-1}~,\\
 &&{\widetilde R}\approx {1\over\sqrt{2}{\widetilde\nu}}
 \left[{\widetilde L}+
 {3\over 2\sqrt{2}}r_0\right]~,\\
 &&\kappa\approx{\sqrt{2}{\widetilde R}^2\over {\widetilde L} r_0}
 \left[1-{3r_0\over 2{\widetilde R}}\right]~,\\
 &&{\widetilde T}={\widetilde M}_P^2{\widetilde\Lambda}
 \approx 2\sqrt{2} M_P^4(r_0 M_P)^{-1}
 \left[1-{3r_0\over 2{\widetilde R}}
 \right]~.
\end{eqnarray}
Note that the entire solution is controlled by one integration constant
${\widetilde \nu}$. In particular, $r_*$ is related to ${\widetilde\nu}$
via
\begin{equation}\label{rstar}
 r_*\approx\epsilon\left\{1+{\widetilde \nu} 
 \left[{{\widetilde L}\over r_0}+{3\over 2\sqrt{2}}\right]^{-1}
 \right\}~.
\end{equation}
Moreover, the solution exists for a continuous range of positive values of the
brane tension also controlled by ${\widetilde \nu}$ via ${\widetilde R}$.  
Note, however, that we have three phenomenological constraints: ${\widetilde
R}$ must be large; ${\widetilde T}$ cannot be smaller than (TeV)$^4$
(supersymmetry breaking considerations); 
the range of values of ${\widetilde T}$ cannot be smaller than 
(TeV)$^4$ (or else this would imply fine-tuning of the brane tension). As
we will see below, we can satisfy these conditions, and this is precisely 
what requires that ${\widetilde \nu}$ be small.

{}Thus, naturalness considerations suggest that (unless the bulk theory
is either very strongly or very weakly coupled)
\begin{equation}
 \beta\equiv r_0 M_P\sim 1~.
\end{equation}
The brane tension can take values in the following window:
\begin{equation}
 {\widetilde T}_{\rm min}< {\widetilde T}<{\widetilde T}_{\rm max}~,
\end{equation}
where
\begin{eqnarray}
 &&{\widetilde T}_{\rm min}\approx {\widetilde T}_{\rm max}-3\sqrt{2}
 {M_P^3\over {\widetilde R}_{\rm min}}~,\\
 &&{\widetilde T}_{\rm max}={2\sqrt{2}\over\beta} M_P^4~,
\end{eqnarray}
where ${\widetilde R}_{\rm min}$ must be somewhat larger then the present
Hubble size. Regardless of the value of ${\widetilde T}_{\rm max}$, the
width of the aforementioned window should not be smaller than (TeV)$^4$ 
(assuming that the supersymmetry breaking scale is around TeV) or else
this would amount to fine-tuning of the brane tension. This then implies
that the 5-dimensional Planck scale $M_P$ cannot be much different from the
4-dimensional Planck scale ${\widetilde M}_P$ (this simply follows from
the fact that we expect ${\widetilde R}_{\rm min}$ to be around $10^{30}~
{\rm mm}$,
while ${\rm TeV}\sim 10^{-15}~{\rm mm}^{-1}$, and ${\widetilde M}_P\sim
10^{15}~{\rm TeV}\sim 10^{30}~{\rm mm}^{-1}$):
\begin{equation}
 M_P\sim {\widetilde M}_P~.
\end{equation}
Note that here various numerical factors might be only roughly
of order 1. Thus, the
window for the brane tension is of oder (TeV)$^4$, but the brane tension
itself is of order ${\widetilde M}_P^4\sim 10^{60}~({\rm TeV})^4$. At first
this might seem as gross fine-tuning. However, this is {\em not} 
necessarily the
case. It is perfectly consistent to have brane tension which is much larger
than (the fourth power of) the supersymmetry breaking scale. Thus, consider a
supersymmetric BPS brane whose tension is or order 
${\widetilde M}_P^4$. Suppose
supersymmetry is dynamically broken on the brane, and the supersymmetry
breaking scale is of order $M_{\rm SUSY}\ll {\widetilde M}_P$. It is clear
that, after supersymmetry breaking (assuming it is spontaneous), the brane
tension remains of oder ${\widetilde M}_P^4$. The contribution to the brane
tension due to the supersymmetry breaking, on the other hand, is
expected to be of order $M_{\rm SUSY}^4$. As we will see in the next
subsection, this is exactly what happens in our solution.

{}Before we end this subsection, let us give the above solution in terms
of $A(r)$ and $B(r)$:
\begin{eqnarray}
 r>\epsilon:~~~&&B(r)=Q(r)~,\\
	       &&A(r)=\ln\left({rQ^\prime(r)\over \epsilon 
 Q^\prime(\epsilon)}\right)~,\\
 r<\epsilon:~~~&&B(r)=\ln\left({\sqrt{2}r_0\over r_*}\right)-\ln\left[1-
 {r^2\over r_*^2}\right]~,\\
	       &&A(r)=\ln\left({{r_*^2+r^2}\over{r_*^2+\epsilon^2}}\right)-
 \ln\left({{r_*^2-r^2}\over{r_*^2-\epsilon^2}}\right)~,
\end{eqnarray}
where $Q(r)$ is the solution to the following differential equation
\begin{equation}
 rQ^\prime=\sqrt{1+{\rm e}^{2Q}-\sqrt{\nu
 \left({{r_*^2+\epsilon^2}\over{r_*^2-\epsilon^2}}\right)^2+{\rm e}^{4Q}}}
\end{equation}
subject to the initial condition
\begin{equation}
 Q(\epsilon)=\ln\left({\sqrt{2}\epsilon~r_*\over{r_*^2-\epsilon^2}}\right)~.
\end{equation}
Note that $r_*$ ($r_*>\epsilon$) 
is not independent but is related to $\nu$ (note that $\nu$ lies in a 
finite interval - see above).
This relation comes from the matching conditions, and its approximate
form is given by (\ref{rstar}). 

\subsection{The BPS Solution}

{}The 5-dimensional bulk action we started with (the Einstein-Hilbert plus
Gauss-Bonnet terms) can be supersymmetrized\footnote{In doing so we will have
fields other than the metric. In the following we will focus on backgrounds
where all bosonic fields other than the metric have vanishing expectation
values. Then the background we presented in the previous subsection is also
a solution to the equations of motion in the supersymmetric case.}. 
We will not give details here, but one can show that for 
${\widetilde \nu}>0$, that is, finite (but large) ${\widetilde R}$ the
solution of the previous subsection does not preserve any supersymmetries.
In particular, there are no non-trivial Killing spinors in this background.

{}However, if we consider the ${\widetilde R}\rightarrow \infty$ limit, we have
a BPS solution. In this limit the brane is flat\footnote{Note that one must
appropriately rescale coordinates in this limit.} and has the geometry of
${\bf R}^{3,1}$. We have only one warp factor $A(z)$, where $z$ is the
coordinate transverse to the brane, and the background metric is conformally
flat:
\begin{equation}
 ds^2=\exp(2A)\eta_{MN}dx^Md x^N~.
\end{equation}
The equations of motion read (note that $z$ here has dimension of length):
\begin{eqnarray}
 &&(A_z)^2\left[1-2\xi (A_z)^2 {\em e}^{-2A}\right]=0~,\\
 &&\left[A_{zz}-(A_z)^2\right]\left[1-4\xi (A_z)^2{\rm e}^{-2A}\right]
 +{1\over 6}{\widetilde \Lambda}{\widetilde L}\delta(z-z_0)=0~,
\end{eqnarray}
where $z_0$ is the location of the brane. As before, let $r_0^2\equiv 4\xi$.
Then we have the following solution (this is the solution corresponding to the
${\widetilde R}\rightarrow \infty$ limit of the solution of the previous
subsection):
\begin{equation}\label{flat}
 A(z)=-\ln\left[{\sqrt{2}\theta(z_0-z)(z_0-z)\over r_0}+1\right]~,
\end{equation}
where $\theta(x)$ is the Heavyside step-function. The matching condition
\begin{equation}
 \left. A_z\left[1-{r_0^2\over 3}(A_z)^2{\rm e}^{-2A}\right]
 \right|^{z_0 +}_{z_0 -} +
 {1\over 6}{\widetilde \Lambda}{\widetilde L}=0
\end{equation}
implies that this solution exists for the following value of the 3-brane
tension:
\begin{eqnarray}
 &&{\widetilde\Lambda}={2\sqrt{2}\over {\widetilde L} r_0}~,\\
 &&{\widetilde T}={\widetilde M_P}^2{\widetilde \Lambda}=
 2\sqrt{2} M_P^3 (r_0 M_P)^{-1}~.
\end{eqnarray}
Note that this is precisely the critical value ${\widetilde T}_{\rm max}$
of the brane tension we found in the previous section (recall that
${\widetilde T}_{\rm max}$ precisely corresponds to the ${\widetilde R}
\rightarrow\infty$ limit).

{}In a bosonic theory the fact that this solution exists only for the
above special value of the brane tension would be interpreted as fine-tuning.
However, in a supersymmetric theory this is not necessarily the case. Indeed,
this value could simply be the BPS saturated value for the brane tension.
In fact, in a supersymmetric version of this model the above solution is
indeed a BPS solution (preserving 1/2 of the supersymmetries). To see this,
let us study Killing spinors in the above background:
\begin{equation}
 {\cal D}_M \varepsilon = 0~.
\end{equation}
Here ${\cal D}_M$ is a generalized covariant derivative:
\begin{equation}
 {\cal D}_M=D_M-{1\over 2} W\Gamma_M~,
\end{equation}
where $D_M$ is the usual covariant derivative containing the spin
connection, and $W$ is interpreted as
the superpotential (we will give its value below). 
The $D$-dimensional gamma matrices $\Gamma_M$ are given by
\begin{equation}
 \Gamma_M=\exp(A){\widehat \Gamma}_M~,
\end{equation}
where ${\widehat \Gamma}_M$ are the corresponding Minkowski gamma matrices
(which are independent of coordinates):
\begin{equation}
 \left\{{\widehat \Gamma_M}~,~{\widehat\Gamma_N}\right\}=2\eta_{MN}~.
\end{equation}
In the following we will use the notation $x^M=(x^m,z)$, and ${\widehat 
\Gamma}_z$ will be the gamma matrix corresponding to the $z$-direction.

{}Thus, we have
\begin{eqnarray}\label{Killing1}
 && 0={\cal D}_z\varepsilon=\partial_z\varepsilon-{1\over 2}W \exp(A){\widehat 
 \Gamma}_z\varepsilon~,\\
 \label{Killing2}
 && 0={\cal D}_m\varepsilon=\partial_m\varepsilon +{1\over 2}
 {\widehat \Gamma}_m\left[A_z{\widehat \Gamma}_z-W\exp(A)\right]
 \varepsilon~.
\end{eqnarray} 
Before solving the Killing spinor equations, let us note that to define an
unbroken supercharge for a given Killing spinor we must make sure that
the global integrability conditions
\begin{equation}
 \left[{\cal D}_M~,~{\cal D}_N\right]\varepsilon=0
\end{equation}
are also satisfied. In the component form these conditions read:
\begin{eqnarray}\label{condi1}
 &&0=\left[{\cal D}_m~,~{\cal D}_n\right]\varepsilon=
 {1\over 4}
 \left[W^2\exp(2A)-(A_z)^2\right]
 \left[{\widehat\Gamma}_m~,~{\widehat\Gamma}_n\right]\varepsilon~,\\
 \label{condi2}
 &&0=\left[{\cal D}_m~,~{\cal D}_z\right]\varepsilon=
 {1\over 2}{\widehat \Gamma}_m \left(\exp(A)~\partial_z W +
 \left[W^2\exp(2A)-A_{zz}\right]{\widehat \Gamma}_z\right)
 \varepsilon~.
\end{eqnarray}
Since in the solution (\ref{flat})
$A_z$ is discontinuous, to satisfy the last condition $W$
must be discontinuous as well. Then only 1/2 of supersymmetries can be 
preserved, and the corresponding Killing spinor has a definite helicity
w.r.t. ${\widehat\Gamma}_z$:
\begin{equation}
 {\widehat\Gamma}_z\varepsilon=\eta\varepsilon~,
\end{equation}
where $\eta$ is either $+1$ or $-1$. We therefore 
have the following BPS equation:
\begin{equation}
 A_z=\eta W\exp(A)~.
\end{equation}
This equation together with the solution (\ref{flat}) then implies
that
\begin{equation}\label{super}
 W={\sqrt{2}\eta\over r_0} \theta(z_0-z)~.
\end{equation}
The Killing spinor is then given by
\begin{equation}
 \varepsilon=\exp\left[{1\over 2}A\right]\varepsilon_0~,
\end{equation}  
where $\epsilon_0$ is a {\em constant} spinor with helicity $\eta$:
\begin{equation}
 {\widehat\Gamma}_z\varepsilon_0=\eta\varepsilon_0~.
\end{equation}
Thus, as we see, 
the solution (\ref{flat}) is a BPS solution preserving 1/2 of supersymmetries.

{}At first it might appear strange that we have non-vanishing superpotential
for $z<z_0$. Note, however, that this is just as well. First of all, the
space for $z<z_0$ is (locally)
AdS$_5$, while for $z>z_0$ it is flat. Secondly, as
was shown in \cite{GB}, the scalar potential in the presence of the 
Gauss-Bonnet term is given by\footnote{This is the 5-dimensional version
of the general $D$-dimensional formula derived in \cite{GB}. In fact, the
formula of \cite{GB} also contains a scalar field, but here $W$ is
independent of any such field.}:
\begin{equation}
 V={3M_P^3\over 2\xi}\left[\left(1-4\xi W^2\right)^2-1\right]~.
\end{equation}
Note that in our setup the bulk scalar potential is vanishing. 
The values of the
superpotential consistent with vanishing $V$ are $W=0$ and 
$W=\pm\sqrt{2}/r_0$, which is precisely what we have in (\ref{super}).

{}Thus, as we see, in the limit where the 3-brane is flat we have a BPS
solution (with ${\cal N}=1$ supersymmetry in four dimensions). On the other
hand, when the brane is a 3-sphere of a large finite radius ${\widetilde R}$
the solution is no longer BPS saturated, rather it is a {\em near}-BPS 
solution. Here we would like to comment on some important properties of
these solutions.

{}Thus, since we have a BPS solution, we can make the discussion 
of the previous subsection 
precise. We can populate the BPS brane with an ${\cal N}=1$ 
supersymmetric theory. Let part of 
this theory contain a chiral superfield $\phi$ with a vanishing tree-level 
superpotential. Let us assume that there is a dynamically generated 
superpotential in this theory. Supersymmetry dictates that the the 3-brane 
action has the following form (here we give only the terms relevant for our 
discussion): 
\begin{equation} 
 S_\Sigma=\int_\Sigma d^4x \left\{\sqrt{-{\widetilde G}}
 \left[{\widetilde M}_P^2 {\widetilde R} 
 -{\widetilde T}_{\rm BPS}-{\widetilde V}(\phi,\phi^*)\right]\right\}~, 
\end{equation} 
where the scalar potential is invariably given by 
\begin{equation} 
 {\widetilde V}(\phi,\phi^*)={\rm e}^{K/{\widehat M}_P^2}\left[ 
 K_{\phi\phi^*}^{-1}\left|F_\phi\right|^2 
 -{3\over {\widehat M}_P^2}{\widetilde W}^2\right]~. 
\end{equation} 
Here $K(\phi,\phi^*)$ is the K{\"a}hler potential, ${\widetilde W}(\phi)$ 
is the superpotential, 
and the F-term is given by: 
\begin{equation} 
 F_\phi={\widetilde W}_\phi+K_\phi {\widetilde W}/{\widehat M}_P^2~. 
\end{equation} 
The scale ${\widehat M}_P$ is the 4-dimensional (reduced) 
Planck scale\footnote{Up to now 
we identified this scale with ${\widetilde M}_P$. However, ${\widehat M}_P$
is not exactly equal ${\widetilde M}_P$. Rather, ${\widehat M}_P^2$ receives
an additional contribution from the higher curvature terms in the bulk as
the second derivatives $A_{zz}$ and $C_{zz}$ have $\delta$-function-like
behavior at the location of the brane. At any rate, in the present context
${\widehat M}_P\sim {\widetilde M}_P$.}. 
Also, note the ${\widetilde T}_{\rm BPS}$ is the critical 
value of the brane tension (that is, 
${\widetilde T}_{\rm BPS}={\widetilde T}_{\rm max}$). 
Note that if ${\widetilde V}$ is non-zero, it contributes 
to the brane tension. 
Here we would like to emphasize that supersymmetry precludes this 
contribution to be of any form 
other than that given above. 

{}Thus, let us assume that at the tree-level ${\widetilde W}=0$. 
Due to the non-renormalization theorem, non-trivial ${\widetilde W}$ 
can only be generated non-perturbatively. Let us assume that
we indeed have such a dynamically generated superpotential. 
Here we would like to show that we can find non-supersymmetric 
solutions (with supersymmetry broken around TeV) such that the 
brane is non-inflating and we do not need any fine-tuning. 
Thus, all 
we need here is to consider a theory where the F-term 
$|F_\phi|\sim ({\rm TeV})^2$ with the scalar potential 
taking a negative value $-{\widetilde V} \sim ({\rm TeV})^4$. 
This, for instance,
can be achieved by taking the minimal K{\"ahler} potential 
$K(\phi,\phi^*)=\phi\phi^*\equiv 
\rho {\widehat M}_P^2$, 
and a non-vanishing 
constant superpotential ${\widetilde W}_\phi=0$. We can obtain 
the latter via, say, gaugino 
condensation\footnote{Here we are assuming that the dilaton 
(as well as any other fields controlling the gauge coupling are stabilized
via, say, the mechanism of \cite{dilaton}.} 
in a non-Abelian supersymmetric gauge theory without matter. 
Let the gauge group $G$ be simple. Then we have 
\begin{equation} 
 {\widetilde W}=h_G\Lambda_G^3~, 
\end{equation} 
where $h_G$ is the dual Coxeter number of the group $G$, and 
$\Lambda_G$ is the dynamically 
generated scale. We then have: 
\begin{equation} 
 {\widetilde V}(\rho)={|{\widetilde W}|^2\over {\widehat M}_P^2} 
 {\rm e}^\rho\left(\rho-3\right)~. 
\end{equation} 
This scalar potential has a non-supersymmetric minimum at $\rho=2$, 
where the F-term 
$|F_\phi|=\sqrt{2}|{\widetilde W}|/{\widehat M}_P$ and the scalar 
potential $V=-e^2 
|{\widetilde W}|^2/{\widehat M}_P^2$. With an appropriate choice of 
$\Lambda_G$ we can have 
the F-term of order (TeV)$^2$ (which implies that the supersymmetry 
breaking scale $M_{\rm SUSY}\sim {\rm TeV}$), and $-V\sim ({\rm TeV})^4$. 

{}Before we end this subsection, we would like to address the following
question. Thus, at first it might appear strange that going from a flat brane
to a spherical one with large radius is consistent with supersymmetry
breaking on the brane around TeV. In particular, note that there are no
covariantly constant spinors on a 3-sphere, so we expect supersymmetry to be
broken, but one might naively expect that, since
the shift in the brane tension should be of order ${\widetilde M}_P^2/
{\widetilde R}^2$ (the curvature of the 3-sphere is of order 
$1/{\widetilde R}^2$), the supersymmetry breaking
scale should be
(at most) of order $\sqrt{{\widetilde M}_P/{\widetilde R}}\sim 1/
{\rm mm}$. Such naive thinking, however, is {\em not} quite correct. Thus, 
note that the brane tension in our solution is given by:
\begin{equation}
 {\widetilde T}({\widetilde R})={\widetilde T}_{\rm max}-3\sqrt{2}{M_P^3
 \over{\widetilde R}}+{\cal O}\left({M_P^2\over {\widetilde R}^2}\right)~,
\end{equation} 
where we are assuming that $1/r_0\sim M_P\sim{\widetilde M}_P$. So we do have
the aforementioned naive contribution of order 
${\widetilde M}_P^2/{\widetilde R}^2$ to the brane tension. However, the
next-to-leading contribution is not this one, rather it is of order
$M_P^3/{\widetilde R}$ (which is related to the fact that we can have
supersymmetry breaking around TeV). The correct way of
thinking about this question is the following. We have a BPS solution with
a flat brane whose tension is of order $M_P^4$. Now we break supersymmetry at
$M_{\rm SUSY}$. The shift in the brane tension is of order $M_{\rm SUSY}^4$.
Nonetheless, in the corresponding non-inflating solution the radius of the
brane, which is now a sphere, is {\em not} of order $M_P/M_{\rm SUSY}^2$
(which would be of order millimeter if we take $M_{\rm SUSY}\sim {\rm TeV}$),
rather it is much larger, namely, of order $M_P^3/M_{\rm SUSY}^4$. And this is
possible due to the fact that the space in the vicinity of the brane is
highly {\em curved}, that is, the warp factor $C(z)$ near the brane is very
large. In turn, this is possible due to the fact that the brane tension is
very large compared with $M_{\rm SUSY}^4$. Put it another way, the effect of
the supersymmetry breaking on the brane is {\em diluted} because of huge
curvature near the brane, so in the resulting non-inflating solution the
radius of the brane is huge!

\section{Comments}

{}In the previous section we constructed a solution where the 3-brane
is a 3-sphere of large radius ${\widetilde R}$ (which is somewhat larger
than the present Hubble size). This solution is completely non-singular, which
is due to the presence of higher curvature terms in the 5-dimensional bulk
action. We chose these terms to be the quadratic (in curvature)
Gauss-Bonnet combination. However, we did this only to make the problem 
computationally tractable. In particular, albeit we do not have a proof of 
this, we expect that solutions with all the qualitative features of our
solution should also 
exist in the presence of generic higher curvature terms in the
bulk. 

{}A key property of our solution is that the 3-brane is non-inflating, and the
solution exists for a continuous range of values of the 3-brane tension.
That is, we do not need any fine-tuning for addressing the cosmological
constant problem in our scenario. Instead, the brane tension is controlled by
an integration constant in this solution. Moreover, we can have large brane
tension. Let us also mention that, 
since our solution does not require fine-tuning,
we should also be able to accommodate small cosmological
constant on the brane, that is, there should also exist solutions
with slowly inflating brane, and such solutions are not expected to require
any fine-tuning either.

{}Since our brane is a non-zero tension 3-sphere, one might worry 
about it collapsing, that is, there might be instability due to a tachyonic
radion mode. Note, however, that since the radius of the 3-sphere is huge,
the (negative) mass squared of this tachyonic mode would be
tiny, and the collapse
would take a very long time. We can estimate this time by noting that the
sphere cannot be shrinking faster than the speed of light. Then if we take
${\widetilde R}$ even a few orders of magnitude larger than the present 
Hubble size, it would take the sphere many times the age of the universe to
shrink.

{}Finally, let us comment on the most important phenomenological 
issue in our scenario. Since in our solution the bulk Planck scale
$M_P$ is of order the brane Planck scale ${\widetilde M}_P$, gravity on the
brane is actually 5-dimensional. Thus, for generic values of $M_P$, we expect
the cross-over scale $r_c\sim {\widetilde L}$ - below $r_c$ gravity is
4-dimensional (that is, if $r_c\gg M_P$), while above $r_c$ gravity is 
5-dimensional. However, in our case $r_c\sim 1/{\widetilde M}_P$, so gravity
is always 5-dimensional. Here we would like to discuss possible ways 
of circumventing this.

{}To begin with, note that we could {\em a priori} take $M_P\ll 
{\widetilde M}_P$ such that the cross-over scale would be of order or somewhat
larger than the present Hubble size thus satisfying phenomenological
constraints \cite{DGZ}. 
However, this would require having $M_P\sim 10-100~{\rm MeV}
\ll M_{\rm SUSY}$. Since the BPS brane tension ${\widetilde T}_{\rm BPS}\sim
M_P^3/r_0$, this would require $r_0\ll 1/M_P$ (this can in principle
be the case if the underlying fundamental theory such as string theory 
is strongly coupled in the bulk) to have the BPS brane tension at least of
order (TeV)$^4$. Note, however, that the window for the brane tension
is of order $M_P^3/{\widetilde R}_{\rm min}$, which would be much smaller than
(TeV)$^4$ unless we allow ${\widetilde R}_{\rm min}$ to be much smaller than
the present Hubble size, namely, ${\widetilde R}_{\rm min}\sim~{\rm mm}$. 
This way we could in principle avoid fine-tuning
the brane tension, but in reality we would be disguising it as another 
fine-tuning. Indeed, for generic contributions to the brane tension coming
from supersymmetry breaking the size of the 3-brane would be of order 
millimeter. To have ${\widetilde R}$ above the present Hubble size, we would
need to fine-tune this contribution with the accuracy of $10^{-60}$.
This approach, therefore, does not solve this issue.

{}To circumvent this, one could consider infinite-volume 
scenarios where a spherical 3-brane is
embedded in higher-than-five dimensional bulk. In the ${\widetilde R}
\rightarrow \infty$ limit we then have a codimension-2 or higher brane.
In the limit of a $\delta$-function-like brane we then do not have
cross-over to higher-dimensional gravity - gravity on the brane is
essentially always 4-dimensional \cite{DG}. More precisely, the cross-over
scale depends on the thickness of the brane and goes to infinity when the
latter goes to zero. However, in this limit the graviton propagator on the
brane is actually singular, and once we include higher derivative terms, 
the cross over scale is expected to be $r_c\sim
{\widetilde M}_P/M_P^2$, which would require $M_P\sim {\rm mm}^{-1}$ 
\cite{DGHS}. More precisely, this is the case when higher derivative
terms are suppressed by powers of $1/M_P$. If, however, we have a greater
suppression factor (this can, for instance, happen if the underlying
fundamental theory is strongly coupled in the bulk), then
$r_c$ could {\em a priori} be much higher. More work needs to be done to
see whether 
non-fine-tuned solutions with 4-dimensional gravity on the brane and large
cross-over scale can be constructed in this context. 
A study in these directions will be reported elsewhere \cite{IK}.

\section{Acknowledgments}

{}Z.K. would like to thank Gia Dvali for useful discussions.

\end{document}